\documentclass[conference]{IEEEtran}
\IEEEoverridecommandlockouts
\usepackage{cite}
\usepackage{amsmath,amssymb,amsfonts}
\usepackage{algorithmic}
\usepackage{graphicx}
\usepackage{textcomp}
\usepackage{xcolor}
\usepackage{hyperref}
\def\BibTeX{{\rm B\kern-.05em{\sc i\kern-.025em b}\kern-.08em
    T\kern-.1667em\lower.7ex\hbox{E}\kern-.125emX}}
\begin{document}

\title{WASD - Water Saving Device}

\author{\IEEEauthorblockN{Luca Magoni}
\IEEEauthorblockA{\textit{Politecnico di Milano} \\
Milan, Italy \\
luca.magoni@mail.polimi.it}
\and
\IEEEauthorblockN{Samuele Marcellino}
\IEEEauthorblockA{\textit{Politecnico di Milano} \\
Milan, Italy \\
samuele.marcellino@mail.polimi.it}
\and
\IEEEauthorblockN{Samuele Rebecchi}
\IEEEauthorblockA{\textit{Politecnico di Milano} \\
Milan, Italy \\
samuele.rebecchi@mail.polimi.it}
}

\maketitle

\begin{abstract}
In response to escalating global drinking water scarcity we propose an innovative, automatic system for reusing clean sink water to flush toilets. Existing solutions for water recycle in the houses involve purifiers and complex treatments, leading to high costs and constant maintenance. WASD, utilizing sensors and a solenoid valve, rapidly detects and separates clean water, directing it to toilet tank while sending non-reusable water to the drain. This cost-effective and user-friendly approach aims to establish sustainable water practices in domestic settings, contributing to solve the shortage of drinking water.
\end{abstract}
\vspace{0.25 cm}
\begin{IEEEkeywords}
Water Quality, Water Monitoring System,Water Saving, Turbidity, pH,  Real Time, Internet of Things, IoT
\end{IEEEkeywords}

\section{Introduction}
Water is one of the most precious and fundamental resources for the survival of all living organisms on Earth. It plays a crucial role in numerous aspects of our daily lives, from personal hygiene to food and energy production. However, despite its undeniable importance, water is a limited resource and increasing population pressures, rapid urbanization and climate change are putting significant stress on the planet’s water resources, leading to a decline in freshwater reserves and worsening water crises in many parts of the world. In this context, it becomes imperative to find methods to prevent waste and reuse water.

Data on water consumption in Italy highlight a concerning reality. According to the latest report from ISTAT [9] on water consumption in Italy, each Italian, on average, receives 236 litres of water per day, but excluding losses in the distribution network, actual consumption is about 174 litres per day. This is still well above the minimum health requirements estimated by the World Health Organization, which range from 50 to 100 litres per day.
The flush of the toilet, with its 10 litres of water per flush, constitutes a significant portion of per capita water consumption, ranging from 20\% to 30\%, equivalent to 34 to 52 litres per day. With each use, this entire volume of potable water is flushed away, even when such a quantity may not be necessary to dispose of waste. Furthermore, the bathroom sink tap has a flow rate of 8 litres per minute. This value is significant when considering that many people leave the tap running while washing their hands or face, needlessly wasting large amounts of water. For example, leaving the tap running while brushing your teeth can result in an additional waste of about 7.5 litres of water [3].

For these reasons, we want to design a new system that is fast, automatic, and non-invasive, allowing clean sink water to be reused for toilet flushing. We aim to design this system to require minimal maintenance and human involvement, placing great emphasis on the accessibility of the product.

The water deriving from baths, sinks, washing machines, and other kitchen appliances is called greywater and, to achieve the objective of water recycling, it is required to respect three criteria (hygienic safety, aesthetics and environmental tolerance) for reuse [14]. However, no international regulations have been published to establish the precise parameters that the water must have to be recycled, also because these parameters can vary a lot depending on the type of reuse.
The reviewing of the published wastewater reuse guidelines indicates that parameters like pH, TSS (total suspended solids), BOD (biochemical oxygen demand), turbidity and coliform shall at least be included for the establishment of a sound grey water reuse guideline. Occasionally, some of the guidelines also contain limits for parameters such as ammonia, phosphors, nitrogen and chlorine residual [10], [8]. 

There are already solutions for water reuse in domestic settings, some of which allow for the reuse of sink water for toilet flushing. However, some of these systems require sedimentation and purifiers [13] resulting slow and requiring regular maintenance, which means additional costs and inconveniences for users. More advanced systems involve water treatments, for example using Membrane BioReactor [12], but these systems are expensive and can only be benefited by large consumers such as hotels, universities, etc.
In contrast, our product does not involve the use of purifiers but utilizes sensors to detect clean water and, with the help of a solenoid valve, separates the flows, directing clean water to the toilet while non-reusable water goes directly into the drain.

\section{Related Works}

\subsection{Real time water quality monitoring systems}

The real-time information about chemical and physical properties of the water is critical in decision making processes, associated with public health safety. 
The traditional method on analysis of water quality parameters that involves the manual collection of samples and laboratory techniques for identifying the contamination [11] results too slow. 
[1] proposes a low cost, and efficient way of water quality monitoring system and controlled usage of water based on IoT (Internet of Things) and WSN (Wireless Sensor Network). This system is divided into two subunits. The first subunit monitors water quality and the second one controls the usage of water at consumer end. The experimental setup consists of temperature sensor, turbidity sensor and pH sensor. It also consists of a solenoid valve which controls water supply. While this system does work as expected, the unit is located in a water tank and, as a result, is not designed to operate with flowing water. Additionally, the system is equipped with few sensors, as it only has detectors for pH, turbidity, and temperature, which allows it to monitor only a limited number of water parameters.

The same issues are encountered in [15] where the proposed system uses four sensors which are pH, turbidity, ultrasonic, DHT-11, microcontroller unit as the main processing module and one data transmission module ESP8266 Wi-Fi module (NodeMCU). The system proposed in this paper is an efficient, inexpensive IoT solution for real-time water quality monitoring. The developed system having Arduino Mega and NodeMCU target boards are interfaced with several sensors successfully. An efficient algorithm is developed in real-time, to track water quality.

[7] has come up with an idea for a low-cost sensor network for real time monitoring and contamination detection in drinking water distribution systems. Their approach is based on the development of low-cost sensor nodes for real-time and in-pipe monitoring and assessment of water quality on the fly. The system used is an Arduino microcontroller with four accommodating sensors: pH, Temperature, Turbidity and Total Dissolved Solids (TDS). Sensors were chosen based on ease of use, measurability (of parameters), portability, as well as being economical and cost-effective. The problem of this system is in the sensors. Data for pH were initially unavailable while the pH sensor was being calibrated. Readings for pH only became available from week 3 onwards after the sensor was readied. Also, the data gathered for pH was inadequate. The turbidity sensor must be improved or replaced altogether. An opening at the top of the turbidity sensor allows water to enter if the sensor is lowered too deep into the stream and the influx causes a direct disturbance in readings. Its short cable meant that it is also severely limited to use at the water edges. Additionally, their turbidity sensor suffered from rust due to seawater exposure and needed to be replaced.

\subsection{Online UV-Vis Spectrophotometers for Water Quality Monitoring}

Online monitoring measures water quality continuously, which allows real-time measurements and process control [4]. It can improve the treatment process with the real-time assessment of both source and treated water quality, identification of contaminants and control of treatment process [5]. It is also useful during the period of rapid changes when quick responses are needed to optimise the process [6].

As studied by Zhining Shi et al. [16] online UV-Vis spectrophotometers have emerged as vital tools in the realm of continuous water quality assessment. They offer the distinct advantage of providing real-time measurements rendering them versatile instruments suitable for a wide array of water quality applications. A brief product review on a submersible UV-Vis spectrophotometer (probe) was conducted in 2006 which summarised the typical applications for wastewater treatment, environmental monitoring and drinking water applications [2]. The landscape of these instruments encompasses various types, each tailored to specific water quality monitoring needs. Single wavelength (SW) devices are adept at measuring specific parameters, such as UV254 or nitrate. Nevertheless, their accuracy can fluctuate, particularly when dealing with source waters characterized by high turbidity. In contrast, full-spectrum instruments proffer in-depth spectra and incorporate built-in algorithms to facilitate the calculation of multiple water quality parameters. However, the precision of these instruments can be contingent upon the unique characteristics of the source water, often necessitating site-specific calibration to ensure measurement accuracy. Advanced UV-Vis instruments utilize proprietary algorithms, grounded in the principles of chemometrics. These algorithms create correlations between UV-Vis spectra and laboratory measurements, thereby offering a means to determine a broad range of water quality parameters. However, it's important to note that the accuracy of these measurements may be source-water dependent, necessitating additional calibration to ensure precise results.

\section{Proposed solution}

\subsection{Systems Specification}

The goal of WASD is to monitor the quality of the water flowing from the sink in real time. Then it divides clean water from waste water, sending the first one to an external tank that can be reused for toilet flushing, while the second one is discarded to the dump. To figure out the needed specifications, both functional and non-functional requirements have to be considered. 
\vspace{0.25 cm}

\textit{Functional requirements:}

\begin{itemize}
    \item \textbf{Water analysis}: The critical phase is sensing. The sensors need to detect the key water properties we are interested in, such as turbidity and pH, to assess its quality.
    \item \textbf{Water separation}: After the water analysis phase, the system must activate the solenoid valve to separate clean water from wastewater.
    \item \textbf{Quick response}: the challenge of this system is to analyse water in real time so that it can detect variations in water quality immediately and not mix clean water with wastewater.
    \item \textbf{Water level monitoring}: Clean water is directed into a tank, and in order to ascertain when the tank reaches its maximum capacity, it is essential to continuously monitor the water level. When the tank is full, the system stops working and it no longer collect clean water but rather discharge it entirely.
\end{itemize}

\textit{Non-functional requirements}:

\begin{itemize}
    \item \textbf{Reliability}: this system works in contact with water and it needs to function efficiently for a long period of time. 
    \item \textbf{Non-intrusiveness}: this system needs to be installed under the sink without make severe changes in the bathroom and without requiring too much space.
    \item \textbf{Low Cost}: since one of the main reasons for installing this system is save on your water bill the cost must remain moderate.
    \item \textbf{Low maintenance cost}: unlike the other solutions present in the market this system aims to require little, if no, maintenance or products reload.
    \item \textbf{Adaptability}: this devise must adapt to different water bodies because different hydric systems provide water with slightly different properties.
\end{itemize}

\subsection{General Architecture}

The Sensing unit is located under the sink, to determine water quality in real time as water flows. A set of two different sensors (pH and turbidity) performs water quality assessment using various sensors integrated with the microcontroller.
Data,coming from sensors, is analyzed by the microcontroller.
The microcontroller processes and analyzes data from the sensing unit, adjusting the operation of the solenoid valve based on the data's compliance with preset parameters. If the tank where clean water is directed becomes full, an ultrasonic sensor detects it, and the microcontroller immediately stops the entire system, allowing all the water to flow into the drain, until the water level decrease and the system reactivates.

\begin{figure}[h]
    \centering
    \includegraphics[width=0.5\textwidth]{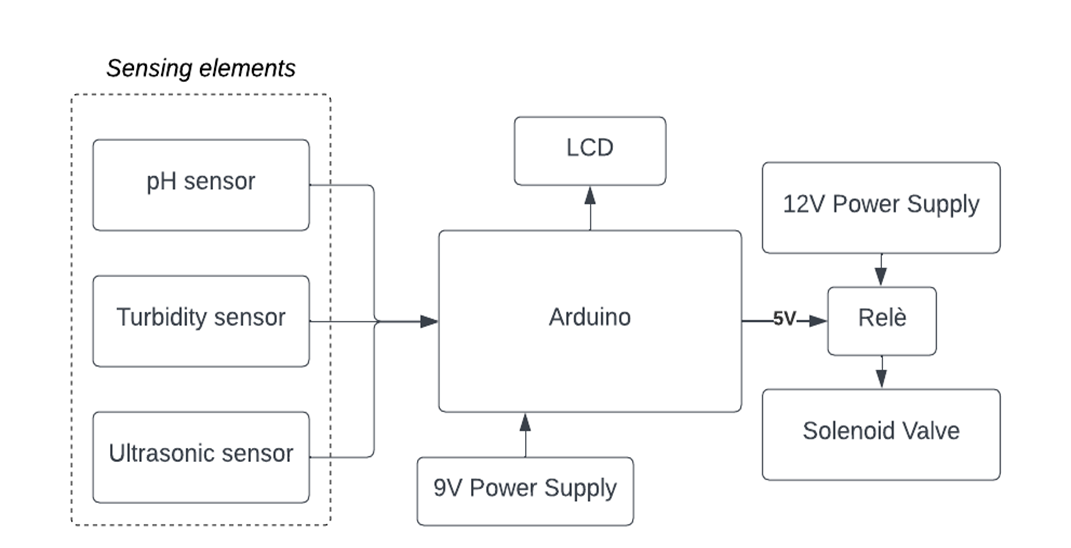}
    \caption{General architecture of the system}
\end{figure}

\begin{enumerate}
    \item \textit{microcontroller}:The water quality monitoring system is designed using \textit{Arduino Uno R3} Microcontroller. Arduino code is written in embedded C programming language using the ARDUINO IDE and compiled, uploaded to the Arduino Uno microcontroller. Arduino handles the processing and analysis of signals from the sensors, as well as the direct control of the solenoid valve.
    \item \textit{pH sensor}: the pH sensor determines the pH value using special electrodes that respond to the presence of hydrogen ions in the solution. When the sensor is immersed in the solution, the electrical potential between the two electrodes changes in response to the concentration of hydrogen ions. This potential variation is then converted into a pH reading, allowing the determination of the acidity or alkalinity level of the solution.
    A pH close to 7 (neutral pH) is often considered ideal for drinking water and many household applications. However, for domestic water use, tolerable pH values are typically within a range of 6.0 to 8.0.
    \item \textit{turbidity sensor}: The turbidity sensor measures how much his light is scattered or absorbed by the suspended particles in the water. The greater the quantity of suspended particles in the water, the more significant the light scattering or absorption. This information is then converted into a turbidity reading, expressed in units of measurement such as NTU (Nephelometric Turbidity Units). The output voltage range represents turbidity value ranging from 0 to 3000 NTU (Nephelometric turbidity unit). Output of circuit is 0$\sim$5V. Turbidity is a measure of water clarity and is determined by the amount of suspended particles present in the water. For domestic water use, turbidity values should generally be below 1 NTU, but for our system we observe that the optimal limit is 50 NTU.
    \item \textit{ultrasonic sensor}: To measure the water level in the tank, we use an ultrasonic sensor, emitting high-frequency sound pulses, beyond the range of human hearing, and then detecting the time it takes for the sound to return after striking an object. Since the speed of sound is well-known, we can derive the distance equivalent to the tank being filled.
    \item \textit{Solenoid Valve}: A solenoid valve is a type of electromechanical device used to control the flow of fluids. It consists of a coil of wire (solenoid) wound around a movable core, typically made of ferrous material. The core is placed inside a hollow coil, and when an electric current is applied to the solenoid, it generates a magnetic field. This magnetic field attracts or repels the movable core, causing it to move within the coil. The movement of the core is linked to a valve mechanism that opens or closes a passage for the fluid to flow. When the solenoid is energized, the valve is open, allowing the fluid to pass through. Conversely, when the solenoid is de-energized, the valve closes, preventing the flow of fluid. A solenoid valve is designed to operate within a specific pressure range. The pressure range within which a solenoid valve functions is a crucial aspect that ensures its compatibility with diverse applications, ranging from low-pressure systems to those with higher pressure demands.
\end{enumerate}

\subsection{System Characteristics}

\textit{Algorithm}:
The three sensors for water quality produce data that need to be processed and analysed by the code in order to compare them with acceptability parameters to determine when to open and close the electro-valve.

\textit{Volt conversion system}:
Since the solenoid valve woks with 12V and Arduino with 5V, to make them interact we add to the system a 5V relay that will be connected to the Arduino, the solenoid valve and a 12V converter; this last one is connected in turn to the home electricity and the electro-valve. In this way, when Arduino will give the signal in 5V, the circuit inside the relay will switch from open to close and a signal in 12V will arrive to the electro-valve. 

\section{Experimental Results}

\subsection{Prototype}
The system is a unified block consisting of three compartments. The initial compartment is dedicated to sensing and incorporates turbidity, pH, and ultrasound sensors. The second compartment manages the water distribution system, comprising a functional electrovalve powered by a 12V transformer and a relay. The final compartment serves as the system's central processing unit, utilizing an Arduino Uno R3 to establish coordination between the first two compartments.
All data collected from the sensor and analyzed by Arduino are displayed on a liquid crystal display.
The model features the components listed in the Table 1.

\begin{table}[htbp]
\caption{Components used for the prototype}
\begin{center}
\begin{tabular}{|c|c|c|}
\hline
\textbf{Component} & \textbf{Model} & \textbf{cost} \\
\hline\hline
Microcontroller & Arduino Uno R3 & €10\\
\hline
turbidity sensor & TSW-30 & €7\\
\hline
pH sensor & Ziqqucu & €11\\
\hline
ultrasonic sensor & HC-SR04 & €2\\
\hline
solenoid valve & AOMAG & €16\\
\hline
\textit{others} & 9V battery, 12V battery, relay & €30 \\
\hline\hline
\multicolumn{2}{|c|}{\textbf{Total}} & \textbf{€76}\\
\hline
\end{tabular}
\label{tab1}
\end{center}
\end{table}

\subsection{Calibration}
In the absence of specific technical documentation, the calibration of sensors became a crucial aspect of ensuring accurate and reliable measurements. We faced with the challenge of calibrating sensors without predefined calibration curves or datasheets, so a methodical and independent approach was necessary. We had to establish a relationship between the measured voltage output from the sensors and the values of the physical parameters being measured.
\begin{itemize}
    \item[-]\textit{turbidity sensor}: In the absence of direct specifications, our approach involved leveraging calibration curves from another turbidity sensor. This empirical correlation derived from the alternate sensor's calibration curves proved to be a viable solution. The results obtained through this unconventional calibration process demonstrated a high degree of reliability, indicating that the relationship between the measured voltage and turbidity values was consistent and reproducible.\\
The Mathematical relationship between voltage from sensor and turbidity value is given by:
\begin{equation}
    Y = -1120.4 V^2 + 5742.3 V - 4352.9
\end{equation}
Where Y is the turbidity value of water in NTU (Nephelometric 
Turbidity unit, 1 NTU = 1 mg/L), and V is the voltage measured by the sensor.
\begin{figure}[h]
    \centering
    \includegraphics[width=0.5\textwidth]{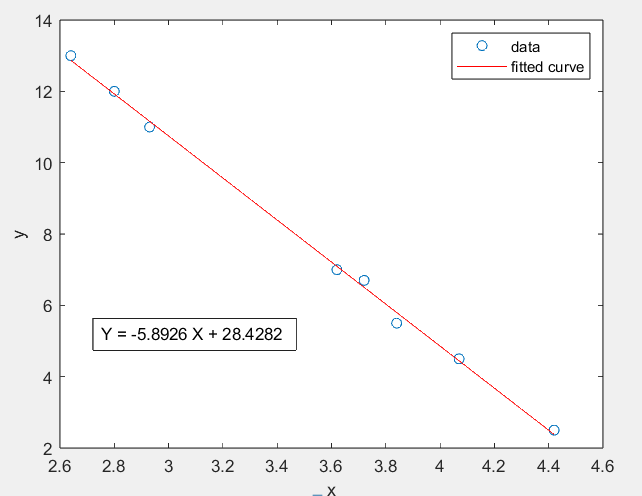}
    \caption{Turbidity calibration curve}
\end{figure}
If the sensor measures voltage values below 2.6V, which is the point where the parabola reaches its maximum, we instruct that the NTU value be capped at 3000 instead of decreasing further.

\item[-]\textit{pH sensor}: the solution involved conducting experimental measurements across a range of substances with known pH values. The sensor's voltage output was measured in various solutions, including hydraulic fluid, natural bleach, universal degreaser, water, milk, coffee, beer, and cola. The pH values for these substances were obtained from product safety data sheets and verified using pH indicator strips. Through meticulous experimentation and data collection, a comprehensive set of voltage-to-pH measurements was obtained.

\begin{table}[htbp]
\caption{Interpolation Data}
\begin{center}
\begin{tabular}{|c|c|c|}
\hline
\textbf{\textit{Solution}}& \textbf{\textit{pH}}& \textbf{\textit{Volt}} \\
\hline\hline
hydraulic fluid & 13 & 2.64\\
\hline
natural bleach & 12 & 2.80\\
\hline
universal degreaser & 11 & 2.93\\
\hline
water & 7 & 3.62\\
\hline
milk & 6.7 & 3.72\\
\hline
coffe & 5.5 & 3.84\\
\hline
beer & 4.5 & 4.07\\
\hline
coca-cola & 2.5 & 4.42\\
\hline
\end{tabular}
\label{tab1}
\end{center}
\end{table}

Utilizing this dataset, an interpolation technique was employed to establish a linear relationship between the measured voltage and pH values. This method allowed us to create a calibration curve that enables the translation of voltage readings from the sensor into accurate and meaningful pH measurements.

\begin{figure}[h]
    \centering
    \includegraphics[width=0.5\textwidth]{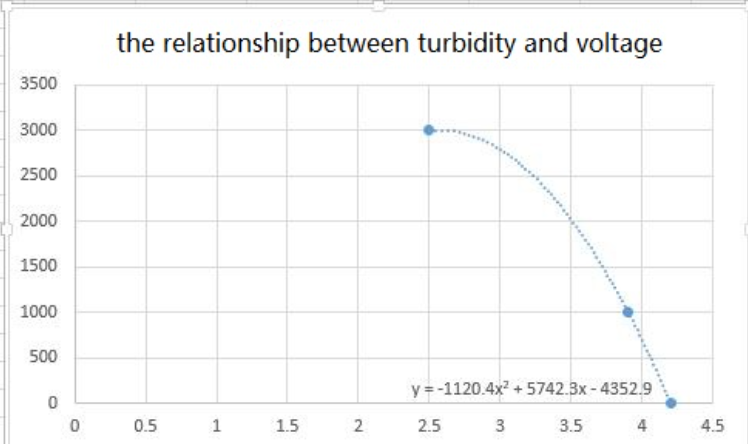}
    \caption{pH interpolation curve}
\end{figure}
\end{itemize}

\subsection{Coding and implementation}
At each loop, the code activates the ultrasonic sensor to continuously detect the water level. The system operates only when the sensor detects that the tank is not full. Initially, a turbidity measurement is taken, and if the required NTU threshold is exceeded, the system skips the pH measurement, deemed unnecessary, and directly closes the solenoid valve. Only if it is detected that turbidity is optimal, the pH measurement is taken. If the pH is also within the correct range (6-8), the solenoid valve is then activated. The code's operation is illustrated by the following flowchart.

\begin{figure}[h]
    \centering
    \includegraphics[width=0.3\textwidth]{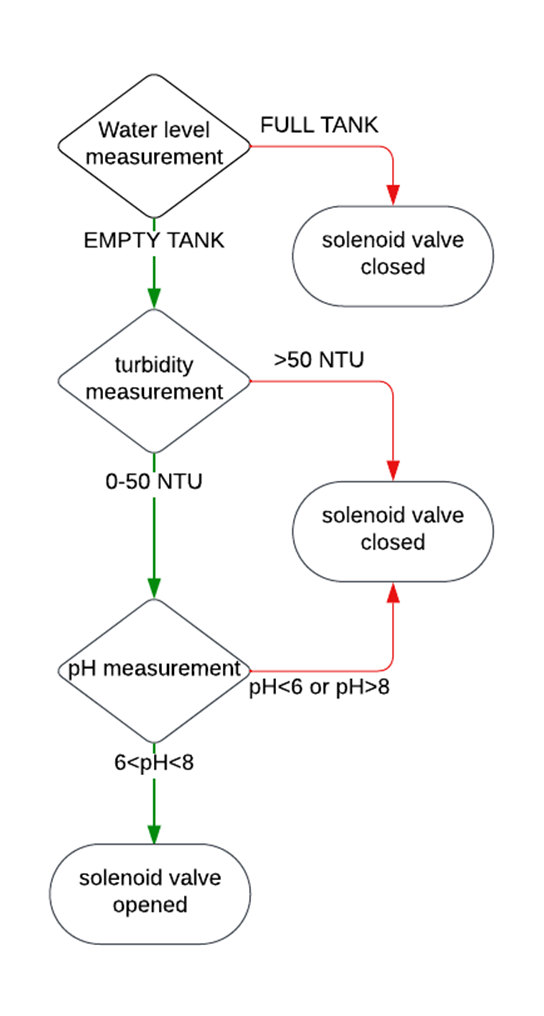}
    \caption{Flowchart of the code at every loop}
\end{figure}

By employing this strategy, the workload on the microcontroller is reduced, avoiding unnecessary measurements and enabling it to perform a greater number of measurements per second. This enhances real-time capabilities.

\subsection{Experiments}

\begin{figure}[h]
    \centering
    \includegraphics[width=0.45\textwidth]{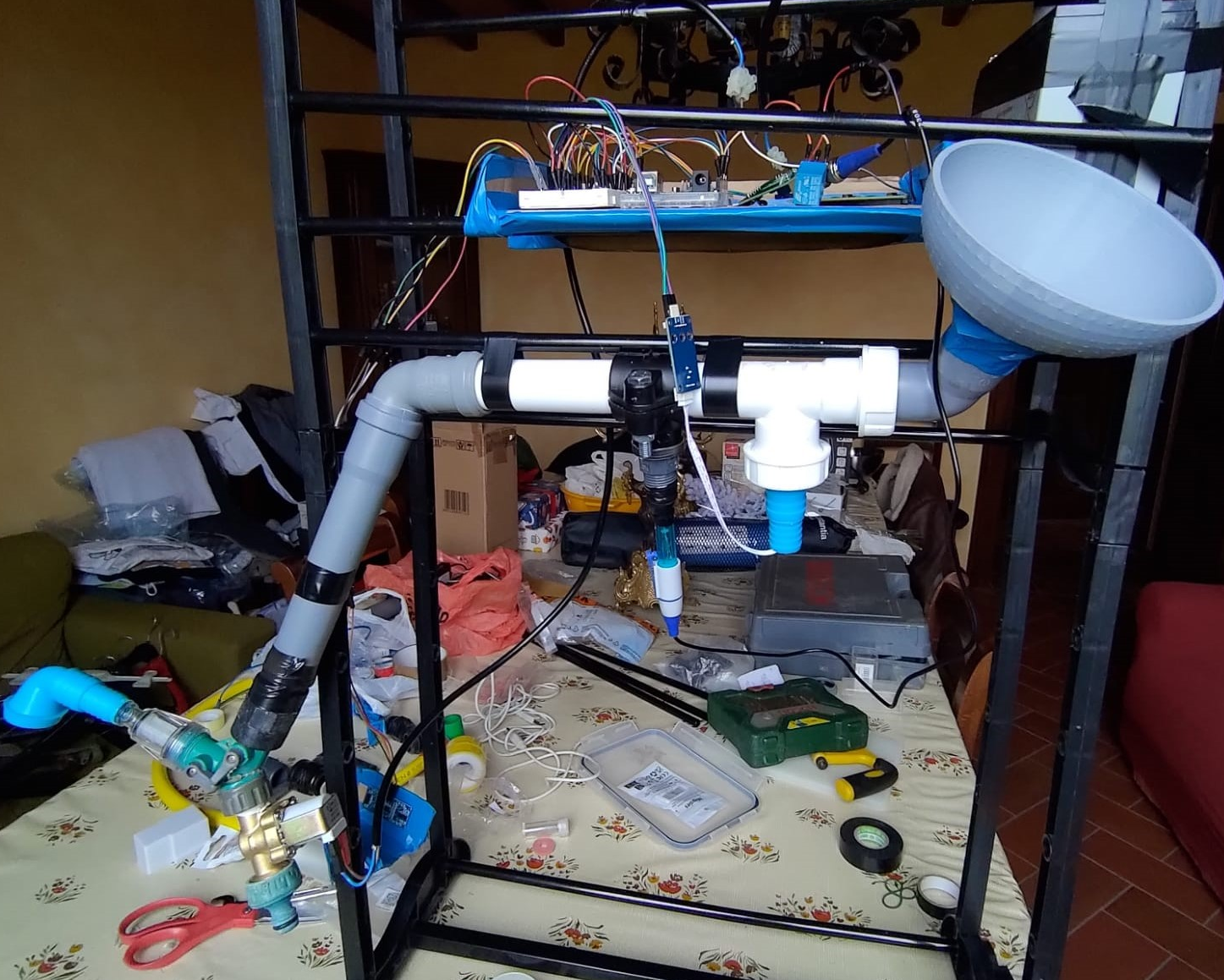}
    \caption{Installed Prototype}
\end{figure}

Once completing the prototype, it is time to introduce same liquid in the system to test his performance.

A problem we suddenly face with clean water is the inconsistency of turbidity data. So, we try to vary the inclination of the sensors which are initially perpendicular to the ground, but we notice that every possible variation produces an even worse reading. We therefore try to vary the inclination of the pipe part in which the sensors are positioned which initially was about 15° downwards. By decreasing the inclination to 0° the turbidity becomes much more stable, almost like in still water, and does not stagnate water in the circuit.

It should also be pointed out that introducing the liquid into the system very violently produces completely incorrect readings. The flow of the liquid must therefore be moderate and regular over time.

We proceed to introduce various substances with different turbidity and pH into the system to check whether these are correctly diverted to the drain or the tank. The results are shown in the table below, where it is possible to read the pH and turbidity values detected by the system Confronting the results with those reported in the literature, the system performed correctly in all measurements. The table shows also that the solenoid valve open or close properly for every substance.

\begin{table}[htbp]
\caption{Experimental Results}
\begin{center}
\begin{tabular}{|c|c|c|c|}
\hline
\textbf{Substance} & \textbf{Turbidity(\textit{NTU})} & \textbf{pH} & \textbf{Solenoid Valve}   \\
\hline\hline
drinking water & 0 & 7.17 & open \\
\hline
Red wine & 3000 & / & Close\\
\hline
Water with little red wine & 0 & 5.51 & Close \\
\hline
Water and bicarbonate & 900-1200 & / & Close \\
\hline
Water and white wine vinegar & 0-40 & 3.10 & Close \\
\hline
Water and Svelto & 600-900 & / & Close \\
\hline
\end{tabular}
\label{tab1}
\end{center}
\end{table}

We also test the consistency of the system with a series of repeated measurements of red wine, all followed by a cleaning of the system with 1.5L of drinking water, thus obtaining the following confusion matrix. The system performs correctly in all measurements by closing the solenoid valve. The system is therefore very consistent.

\begin{table}[htbp]
\caption{Confusion Matrix}
\begin{center}
\begin{tabular}{|c|c|c|}
\hline
Samples: 10 & \textbf{Open valve} & \textbf{Closed valve}\\
\hline\hline
\textbf{water} & 0 & 0 \\
\hline
\textbf{wastewater} & 0 & 100\%\\
\hline
\end{tabular}
\label{tab1}
\end{center}
\end{table}

One fact that we expected to be among the most challenging is detecting the quality of the water in a sufficiently short time to make this system perform well.

A countermeasure we take is to write in the code that the solenoid valve remains closed, thus channelling the liquids into the drain, until the first reading of both sensors is completed. In this way, if the speed of the liquid is such that it travels along the entire tube before the first reading, not a drop of unknown liquid can end up in the tank.

By carrying out experiments with substances with different values of turbidity we realize that the variation is read almost instantaneously, with usually a couple of seconds of delay necessary to eliminate any residues of previous liquid in the tube.

By using substances with excellent turbidity and different pH, the data is no longer always so good. If we let water flow through the circuit and then replace it with a substance with a non-neutral pH, such as white wine vinegar, we obtain a rapid decrease in the pH value read with a consequent immediate closing of the solenoid valve. However, if we then reintroduce water into the system, we need to pour in at least 1L before getting a good reading again. 

This means that the pH sensor takes a long time to return reading correctly after introducing a particularly acidic or basic substance.

We also mention the following little complications that can be easily fixed with a slightly more expensive equipment:
\begin{itemize}
    \item If the system remains off for a few hours the sensors need to be slightly recalibrated.
    \item The solenoid valve has a hole that is too narrow so even when it is open some water goes into the drain.
\end{itemize}
All the system is functional and with small adjustments this prototype can actually be used on an industrial scale.

\section{CONCLUSIONS}
WASD is functional for controlling the quality of a moving liquid and its sorting and deserves to be further studied. Before it is ready to be produced on an industrial scale, further tests are necessary with more sophisticated and expensive sensors that can further reduce reading times and that can last over time without the need for constant recalibration and maintenance. 

It is possible to get an excellent set of sensors by spending from 100 to 300 euros and this makes this system very convenient also on an economic level. In fact, the experiments show that with this system is possible to recover up to 52L per day of greywater as initially hypothesized.

Possible future developments are to be found in the reduction in size and portability of the product. Since it is necessary to intervene on the plumbing system, WASD can only be installed in bathrooms at the time of construction, or during a restoration. If we could create a portable system that can be easily installed in the bathroom without major plumbing interventions, this system could be applied in every home and have a tangible impact on global water consumption in the very near future.

\bibliographystyle{plain}
\bibliography{bibliography}

\section{REFERENCE}
\small
[1] Achyutprasad Annigeri and Sujatha B. Optimized domestic water quality and consumption monitoring system based on iot. In 2021 IEEE Mysore Sub Section International Conference (MysuruCon), pages 1–5, 2021.

[2] Joep van Den Broeke, Albert Brandt, Andreas Weingartner, and Franz Hofst¨adter. Monitoring of organic micro contaminants in drinking water using a submersible uv/vis spectrophotometer. In Jaroslav Pollert and Bozidar Dedus, editors, Security of Water Supply Systems: from Source to Tap, pages 19–29, Dordrecht, 2006. Springer Netherlands.

[3] Gruppo CAP. Consumo di acqua: qual `e la media giornaliera. 2023. 

[4] AG Capodaglio and A Callegari. Water supply systems security: novel technologies for the online monitoring of unforeseeable events. WIT Trans. Built Environ. Saf. Secur. Eng, 51:251–263, 2015.

[5] Claudia Cascone. Optical sensors in drinking water production: Towards automated process control in relation to natural organic matter. Acta Universitatis Agriculturae Sueciae, (2021: 17), 2021.

[6] Chris Chow, Chris Saint, Luke Zappia, Rita Henderson, Gareth Roeszler, Rob Dexter, Tung Nguyen, Richard Stuetz, Amanda Byrne, Rino Trolio, Jeremy Lucas, Sally Williamson, and Simon Wilson. Online water quality monitoring: The voice of experience. Water: Journal of the Australian Water Association, 41(2):60–62, 65–66, 68, 2014.

[7] Wong Jun Hong, Norazanita Shamsuddin, Emeroylariffion Abas, Rosyzie Anna Apong, Zarifi Masri, Hazwani Suhaimi, Stefan Herwig G¨odeke, and Muhammad Nafi Aqmal Noh. Water quality monitoring with arduino based sensors. Environments, 8(1), 2021.

[8] A.C. Hurlimann and J.M. McKay. What attributes of recycled water make it fit for residential purposes? the mawson lakes experience. Desalination, 187(1):167–177, 2006. Integrated Concepts in Water Recycling.

[9] ISTAT. Va perduto oltre un terzo dell’acqua immessa nella rete di distribuzione. 2022.

[10] P. Jeffrey and B. Jefferson. Public receptivity regarding “in-house” water recycling: results from a UK survey. Water Supply, 3(3):109–116, 06 2003.
[11] O. Korostynska, A. Mason, and A. I. Al-Shamma’a. Monitoring Pollutants in Wastewater: Traditional Lab Based versus Modern Real-Time Approaches, pages 1–24. Springer Berlin Heidelberg, Berlin, Heidelberg, 2013.

[12] Nir Liberman, Semion Shandalov, Chaim Forgacs, Gideon Oron, and Asher Brenner. Use of mbr to sustain active biomass for treatment of low organic load grey water. Clean Technologies and Environmental Policy, 18, 04 2016.
[13] J.G March, M Gual, and F Orozco. Experiences on greywater re-use for toilet flushing in a hotel (mallorca island, spain). Desalination, 164(3):241–247, 2004.

[14] Erwin Nolde. Grey water reuse systems for toilet flushing in multi-storey buildings – over ten years experience in berlin. Urban Water, 1:275–284, 12 2000

[15] Sathish Pasika and Sai Teja Gandla. Smart water quality monitoring system with cost-effective using iot. Heliyon, 6(7), 2020.

[16] Zhining Shi, Christopher W. K. Chow, Rolando Fabris, Jixue Liu, and Bo Jin. Applications of online uv-vis spectrophotometer for drinking water quality monitoring and process control: A review. Sensors, 22(8), 2022.

\end{document}